\documentclass[pdflatex,sn-apa]{sn-jnl}

\makeatletter
\@twosidefalse
\@mparswitchfalse
\makeatother

\usepackage{graphicx}%
\usepackage{multirow}%
\usepackage{amsmath,amssymb,amsfonts}%
\usepackage{amsthm}%
\usepackage{mathrsfs}%
\usepackage[title]{appendix}%
\usepackage{xcolor}%
\usepackage{textcomp}%
\usepackage{manyfoot}%
\usepackage{booktabs}%
\usepackage{algorithm}%
\usepackage{algorithmicx}%
\usepackage{algpseudocode}%
\usepackage{listings}%

\usepackage{tabularx}
\usepackage{booktabs}    
\usepackage{array}
\usepackage{rotating}
\usepackage{float}
\usepackage{lscape}
\usepackage{hvfloat}

\PassOptionsToPackage{hyphens}{url}
\usepackage{hyperref}

\usepackage{etoolbox}
\apptocmd{\sloppy}{\hbadness 10000\relax}{}{}
\theoremstyle{thmstyleone}%
\theoremstyle{thmstyletwo}%

\theoremstyle{thmstylethree}%

\begin{document}

\title[Article Title]{aiPlato: A Novel AI Tutoring and Stepwise Feedback System for Physics Homework}

%%=============================================================%%
%% GivenName	-> \fnm{Joergen W.}
%% Particle	-> \spfx{van der} -> surname prefix
%% FamilyName	-> \sur{Ploeg}
%% Suffix	-> \sfx{IV}
%% \author*[1,2]{\fnm{Joergen W.} \spfx{van der} \sur{Ploeg} 
%%  \sfx{IV}}\email{iauthor@gmail.com}
%%=============================================================%%

\author*[1, 2]{\fnm{Atharva} \sur{Dange}}\email{dange98@mit.edu}

\author[1]{\fnm{Ramon E.} \sur{Lopez}}\email{relopez@uta.edu}

\affil[1]{\orgdiv{Physics}, \orgname{University of Texas at Arlington}, \orgaddress{\city{Arlington}, \postcode{76019}, \state{Texas}, \country{USA}}}

\affil[2]{\orgdiv{Physics}, \orgname{Massachusetts Institute of Technology}, \orgaddress{\city{Cambridge}, \postcode{02139}, \state{Massachusetts}, \country{USA}}}

\author[3]{\fnm{Nimish} \sur{Shah}}\email{nimish@aiplato.ai}

\author[3,4]{\fnm{Louis} \sur{Deslauriers}}\email{louisdeslauriers@fas.harvard.edu}

\affil[3]{\orgdiv{}, \orgname{aiPlato}, \orgaddress{\city{San Francisco}, \postcode{94501}, \state{California}, \country{USA}}}

\affil[4]{\orgdiv{Physics}, \orgname{Harvard University}, \orgaddress{\city{Cambridge}, \postcode{02138}, \state{Massachusetts}, \country{USA}}}

%%==================================%%
%% Sample for unstructured abstract %%
%%==================================%%

\abstract{This exploratory study examines the classroom deployment of aiPlato, an AI-enabled homework platform, in a large introductory physics course at the University of Texas at Arlington. Designed to support open-ended problem solving, aiPlato provides stepwise feedback and iterative guidance through tools such as \textit{Evaluate My Work} and \textit{AI Tutor Chat}, while preserving opportunities for productive struggle. Over four optional extra-credit assignments, the platform captured detailed student interaction data, which were analyzed alongside course performance and end-of-semester survey responses. We examine how students engaged with different feedback tools, whether engagement patterns were associated with performance on the cumulative final exam, and how students perceived the platform’s usability and learning value. Students who engaged more frequently with aiPlato tended to achieve higher final exam scores, with a mean difference corresponding to a standardized effect size of approximately 0.81 between high- and low-engagement groups after controlling for prior academic performance. Usage patterns and survey responses indicate that students primarily relied on iterative, formative feedback rather than solution-revealing assistance. As a quasi-experimental pilot study, these findings do not establish causality and may reflect self-selection effects. Nonetheless, the results demonstrate the feasibility of integrating AI-mediated, stepwise feedback into authentic physics homework and motivate future controlled studies of AI-assisted tutoring systems.}

\keywords{AI in Education, Intelligent Tutoring Systems (ITS), Homework Platfroms, Physics Education, LLMs}

%%\pacs[JEL Classification]{D8, H51}

%%\pacs[MSC Classification]{35A01, 65L10, 65L12, 65L20, 65L70}

\maketitle

\section{Introduction}\label{sec1}

Homework has long served as a cornerstone of STEM education, particularly in physics \citep{bempechatMotivationalBenefitsHomework2004, konturBenefitsCompletingHomework2013, bonhamOnlineHomeworkDoes2001}, where it reinforces conceptual understanding and cultivates multi-step problem-solving skills \citep{bempechatMotivationalBenefitsHomework2004}. Well-designed assignments help students build fluency with equations \citep{vasuki2016curriculum}, develop persistence through cognitive struggle, and transfer classroom learning into contextual problems \citep{dominguezIntegrationPhysicsMathematics2023}. Over the past two decades, digital homework platforms such as WebAssign, WileyPLUS, and Expert TA have transformed the logistics of homework delivery by enabling scalable grading, instant feedback, and standardized assessments in large-enrollment courses \citep{kuoCaseStudiesMyMathLab2012, basitere2017evaluation}.

However, despite these operational improvements, many online homework systems fall short of supporting deep learning. Feedback is typically correctness-based, either binary (right or wrong) or limited to generic hints, offering little insight into the student’s reasoning process or misconceptions \citep{gladdingClinicalStudyStudent2015, buhr202021st}. Students often develop surface-level strategies, such as answer guessing or brute-force substitution, while instructors remain in the dark about how students are thinking, where they get stuck, or what conceptual gaps persist \citep{wu2020undergraduate}. These limitations are especially pronounced for open-ended or symbolic problems common in physics, where understanding the "how" is just as critical as the "what."

In response to these challenges, recent advances in AI-powered tutoring systems offer new possibilities for improving feedback, personalization, and formative assessment at scale \citep{alshaikhAIMachineLearning2021, bhatiaLeveragingAITransform2024, linArtificialIntelligenceIntelligent2023, rizviInvestigatingAIPoweredTutoring2023}. This paper examines one such platform, aiPlato \citep{aiplato}, which is designed to provide step-by-step, context-sensitive feedback on free-response physics problems. Drawing from a pilot deployment in a large introductory mechanics course at the University of Texas at Arlington, we explore how students interact with aiPlato’s feedback tools, whether greater engagement translates to improved learning outcomes, and how instructors and students perceive its value. The full statistical analyses and learning outcome results are presented in section \ref{sec4}, and this paper situates the study within the broader context of intelligent tutoring systems, LLM-based tutors, and the evolving role of AI in education.

\section{Intelligent Tutoring Systems (ITS) in Physics Education}\label{sec2}

Intelligent Tutoring Systems (ITS) represent a major milestone in the evolution of educational technology, offering real-time, adaptive support that mimics the responsiveness of a human tutor \citep{nwanaIntelligentTutoringSystems1990, graesserIntelligentTutoringSystems2012, vanlehn2005andes}. In physics education, where conceptual reasoning and multi-step problem-solving are critical, ITS aim to diagnose student misconceptions, guide learners through complex tasks, and provide formative feedback during the problem-solving process itself \citep{albaceteConceptualHelperIntelligent2000, myneniInteractiveIntelligentLearning2013}.

Early ITS implementations in mathematics and computer science demonstrated strong potential for improving learning outcomes through personalized instruction. Systems like Andes \citep{schulzeAndesIntelligentTutor2000}, one of the earliest ITS developed for classical physics, sought to support students by interpreting symbolic input, checking for logical consistency, and providing stepwise hints without immediately revealing answers \citep{akyuzEffectsIntelligentTutoring2020}. These systems used rule-based approaches grounded in domain-specific knowledge, enabling them to guide students through canonical physics problem types.

However, widespread adoption of ITS in physics has been challenging, in part to the inherent complexity of modeling open-ended, algebraically rich reasoning pathways that are common in physics problem solving \citep{fengSystematicReviewLiterature2021}. Unlike multiple-choice assessments, physics problems often allow for multiple valid approaches, varied intermediate steps, and symbolic answers - all of which present challenges for rigid or rule-based systems.

Despite these challenges, research shows that when properly designed, ITS can deliver learning gains comparable to those achieved through one-on-one human tutoring \citep{vanlehnRelativeEffectivenessHuman2011}. In particular, ITS that provide timely, specific, and process-oriented feedback, and not just correctness judgments, can help students identify misconceptions, persist through confusion, and reflect on their reasoning \citep{merrillEffectiveTutoringTechniques1992}. These systems embody the principles of formative assessment, emphasizing the reasoning behind student errors.

Recent efforts have extended ITS capabilities by incorporating mastery-based progression \citep{fangMetaanalysisEffectivenessALEKS2019} and simulation-driven environments \citep{wiemanTeachingPhysicsUsing2010}. In these systems, students must demonstrate conceptual understanding before advancing, while being encouraged to plan, test, and revise their approaches. Studies show that this type of scaffolding improves conceptual retention, especially when combined with feedback that adapts to individual learning trajectories \citep{changEffectivenessScaffoldingWebBased2009}.

Still, most ITS in physics remain limited in scope, constrained by the difficulty of encoding every possible reasoning pathway and response structure. As a result, their use has largely been confined to structured problems within tightly defined content areas. The growing interest in more flexible, AI-powered feedback mechanisms, particularly those leveraging natural language processing and data-driven analytics, reflects a desire to overcome these limitations.

\subsection{LLM-Based Tutors}\label{subsec2.1}

Recent advances in large language models (LLMs), including those based on transformer architectures like GPT-4, have led to widespread interest in their use as educational tools \citep{jiangAnswersLargeLanguage2024, wangLargeLanguageModels2024}. These models are capable of producing human-like responses in natural language and can be deployed in a wide range of contexts, including tutoring in technical subjects such as physics \citep{scarlatosTrainingLLMbasedTutors2025, razafinirinaPedagogicalAlignmentLarge2024}. Instructors and developers have been exploring how LLMs might assist students by providing real-time, conversational feedback that is scalable and widely accessible.

One commonly cited benefit of LLM-based tutors is their ability to respond instantly to diverse student questions. These systems can adjust their responses based on student inputs and offer individualized feedback that is not limited by time or classroom constraints \citep{thomasLLMGeneratedFeedbackSupports2025}. As a result, learners with different levels of background knowledge or varying schedules can receive help when they need it, which makes these systems attractive in asynchronous or remote learning environments \citep{huGenerativeAIEducation2025}.

LLM tutors also offer potential advantages in terms of scalability. Unlike human instructors or teaching assistants, who are limited by time and availability, these models can respond to thousands of queries at once without additional staffing. Instructors may also benefit from using LLMs to automate repetitive tasks such as drafting feedback or generating quiz content \citep{dangeUSINGGENERATIVEAI2025}, saving time that could be redirected toward higher-impact instructional design.

The interactive format of LLM-based tutors is another strength. Through a conversational interface, students can ask questions, revise their understanding, and explore alternate solution paths in a low-pressure environment. This type of exploratory engagement can increase motivation and foster greater participation in problem-solving activities \citep{aslanEarlyInvestigationCollaborative2025}.

\begin{figure}[h]
\centering
\includegraphics[width=\textwidth]{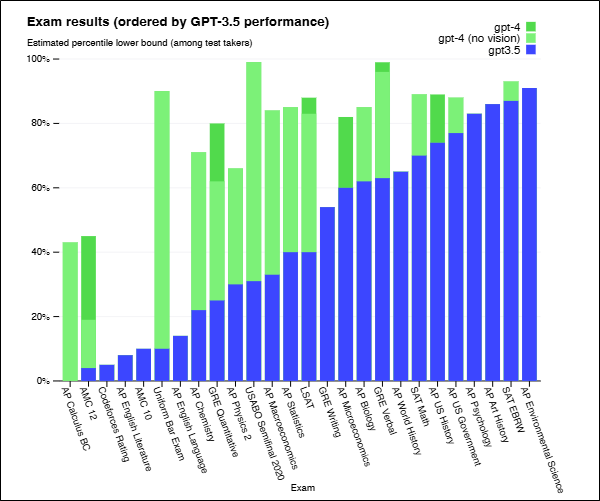}
\vspace{-1em} % <-- optional if caption is too far
\caption{Performance comparison of GPT-3.5 and GPT-4 (with and without vision) across standardized exams. Notably, GPT-4 scored in the 43rd percentile on AP Calculus BC (column 1) and the 66th percentile on AP Physics 2 (column 10) \citep{openaiGPT4TechnicalReport2024}}
\label{Fig. 1}
\end{figure}

However, these strengths are accompanied by notable limitations, particularly in subjects like physics that require stepwise reasoning and technical precision, see Fig. \ref{Fig. 1}. One of the most serious concerns is the tendency of LLMs to produce responses that sound correct but contain inaccurate or misleading information. This issue, often described as hallucination, arises when the model generates plausible content that is not based on verified knowledge \citep{hoMitigatingHallucinationsLarge2024, huangSurveyHallucinationLarge2025}. In physics, where small errors in logic or numerical accuracy can lead to major conceptual misunderstandings, this risk is especially problematic \citep{elsayed2024impact}.

Another common challenge is that LLMs often skip intermediate reasoning steps or fail to clearly explain how an answer is derived. Instead of identifying a student's specific error and offering targeted guidance, the model may return a complete solution without addressing the underlying confusion \citep{hahnSystematicReviewEffects2021}. In physics education, where learning is often rooted in building a clear understanding of the process rather than just arriving at a final result, this lack of structure can make the feedback less useful.

There are also concerns about the consistency and reliability of feedback. Because LLMs are trained on general data rather than structured domain-specific knowledge \citep{yangEmpowerLargeLanguage2023}, the quality of responses can vary widely. Instructors and students may encounter responses that overlook key ideas, contain ambiguous reasoning, or offer suggestions that are difficult to apply to the original problem \citep{debuseBenefitsDrawbacksComputerbased2016}.

Students may also come to rely on LLMs for quick answers rather than using them to guide their thinking. If the model provides full solutions too easily, students might bypass productive struggle and develop superficial understanding \citep{zhaiEffectsOverrelianceAI2024}. In some cases, they may even accept incorrect answers without checking, especially if the explanation appears fluent and confident \citep{grahamImpactChatGPTReliance2025}.

Training biases are additional concerns: because LLMs reflect the data they were trained on, they may unintentionally reinforce language patterns or content that disadvantage themselves (and their users). For instance, popular image generation repeated failures to generate wine glasses filled to the brim \citep{Muratgul_2025}, or clock faces overwhelmingly display the time as 10:10 (a common marketing pose in commercial imagery \citep{karimWhy10102017}) rather than representing a diverse range of times. These examples illustrate how AI models internalize and reproduce the most frequently encountered patterns from their datasets, regardless of whether those patterns reflect the full diversity of real-world contexts. Another example, students who are not native English speakers or who come from different cultural or academic backgrounds may find the feedback less intuitive or useful \citep{isomidin2025using}.

More recent research has also raised questions about how these models behave when fine-tuned on educational data. In some cases, LLMs that are exposed to incorrect student reasoning during training may begin to mimic those patterns, reducing their reliability as instructional tools \citep{sonkarStudentDataParadox2024}. While technical strategies such as prompt engineering or Application programming interface (API) integration \citep{Auger2024} may help address some of these problems, they are often difficult to implement in real classroom settings without technical expertise.

Given these limitations, LLM-based tutoring must be approached carefully. The flexibility and accessibility of these systems offer real value, but their effectiveness depends on how well they support structured reasoning, identify misconceptions, and offer reliable feedback in real time. These limitations have motivated the development of new approaches that aim to retain the strengths of conversational AI while addressing its weaknesses in precision and pedagogy. While LLM-based tutors offer valuable benefits in terms of personalization, accessibility, and engagement, their current limitations in accuracy, reasoning, and feedback consistency present significant challenges in domains that demand precision and structure. Understanding these limitations, while preserving the strengths of conversational AI, opens the door to designing new tools that can support authentic learning in physics. Such tools must not only guide students through complex reasoning tasks but also provide instructors with reliable, easy-to-use systems that align with course goals and instructional workflows. 

\subsection{Need for a Novel AI Platform for Physics Learning}\label{subsec2.2}

In response to the limitations of both traditional ITS and LLM-based tutors, a new category of systems has begun to emerge, what the authors like to call: Artificial Intelligent Tutoring Systems (AITS). These platforms combine the structured scaffolding of ITS with the adaptability, personalization, and real-time feedback enabled by artificial intelligence \citep{akintolaAdaptiveAISystems2025}. AITS are designed to support domain-specific reasoning, guide students through multi-step problem solving, and provide instructors with meaningful learning analytics \citep{tanArtificialIntelligenceenabledAdaptive2025}.
Unlike correctness focused platforms, AITS are centered on conceptual understanding. Rather than simply confirming whether an answer is right or wrong, they strive to highlight misconceptions, acknowledge partial progress, and offer guidance tailored to the student's approach. This shift from answer validation to process-oriented support is critical for promoting deeper, more durable learning. AITS also provide feedback during the actual process of problem solving, helping students reflect and revise in real time \citep{vermuntProcessorientedInstructionLearning1995}.

A distinguishing feature of AITS is their focus on productive struggle \citep{murdochFeelingHeardInclusive2020}. These systems are designed to help students engage with challenging problems while receiving feedback that addresses the root causes of their mistakes \citep{granbergDiscoveringAddressingErrors2016}. Compared to general-purpose LLM tutors, AITS offer more accurate, structured, and context-sensitive feedback that aligns with instructional goals. They are grounded in expert verified knowledge and built to interpret symbolic and multi-step reasoning in ways that probabilistic text generation cannot consistently replicate.

This paper presents a pilot study of one such AITS: aiPlato, developed for physics education. Deployed in an introductory mechanics course at the University of Texas at Arlington, aiPlato was used for four extra credit assignments involving open-ended problems, primarily derived from OpenStax \citep{staffordOpenStax2018}. This study investigates how students interacted with aiPlato, whether that engagement supported learning, and how students perceived the system. This study was guided by three central research questions:

\begin{enumerate}
  \item How do students engage with aiPlato’s feedback tools, and what kinds of insights do their interactions provide to instructors about student thinking and learning progress?
  \item Is there a measurable relationship between student engagement with aiPlato and their performance on the cumulative final exam?
  \item What are students’ perceptions of aiPlato’s usefulness, usability, and learning value, based on end-of-semester survey responses and qualitative feedback?
\end{enumerate}

\section{aiPlato:  A Walkthrough}\label{sec3}

This section provides an illustrative walkthrough of how a student interacts with the aiPlato platform while solving a representative physics problem. Rather than generalizing about the platform’s capabilities in abstract terms, we focus on a single example drawn from the aiPlato problem corpus to describe how various features are designed to support student reasoning. The goal is to familiarize the reader with the problem-solving workflow and tools available within the interface, as they are typically encountered during assignment use. 

The following Fig. \ref{Fig. 2} shows the general layout of the aiPlato interface as experienced by students during problem-solving sessions. This workspace integrates the problem statement, input area, equation tracking panel, and optional feedback tools into a single screen. Students interact with this interface while constructing their responses, requesting support, or submitting final answers. In the example used here, students were asked to derive an expression for the gravitational force acting on a mass at a distance r<R from the center of the planet. 

Next, as Fig. \ref{Fig. 3} shows, students began their solution using either typed, handwritten input, or a combination of the two in the Workspace Grid. The system parsed the scanned content and extracted line-by-line equations into the right-hand Equation Board. After Evaluate My Work, the student received targeted feedback identifying a conceptual error in Line 3 and 4, where surface area was mistakenly used instead of volume. The system provided not only corrective comments but also context-specific hints to guide revision without revealing the final answer. In addition to qualitative feedback, the platform automatically assigned partial credit to each step based on the logic and completeness of the derivation. Other research’s efforts on awarding partial credit via a mobile app \citep{hsiaoMobileGradingPaperBased2016} or via Microsoft Excel \citep{carberryBringingNuanceAutomated2019} can be convoluted, come with a learning curve or impossible for open-ended questions. This form of grading allowed students to receive credit not just for final answers, but also for intermediate reasoning, including steps that may not have been explicitly shown but were inferable based on the context. The total score reflected both the explicit input and the system’s interpretation of the student's conceptual progression.

\begin{figure}[h]
\centering
\includegraphics[width=0.8\textwidth]{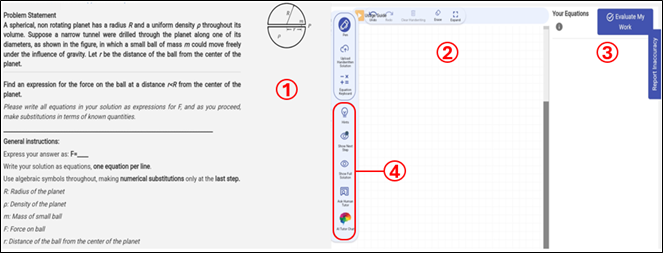}
%\vspace{-1em} % <-- optional if caption is too far
\caption{General layout of the aiPlato platform as seen by students during assignments.}
\label{Fig. 2}
\begin{flushleft}
\begin{enumerate}
\small \item \textbf{Problem Statement Area}: Displays the physics problem, expected final answer format, variable definitions, and supporting diagram.\\ 
\small \item \textbf{Workpace Grid}: Allows students to handwrite or type equations and sketches using a stylus, touchscreen, or mouse. \\
\small \item \textbf{Equation Board}: Records submitted steps line-by-line, supports editing and evaluation via \textit{Evaluate My Work} button.\\
\small \item \textbf{Support Tool Sidebar}: Contains optional features including \textit{Hints}, \textit{Show Next Step}, \textit{Show Full Solution}, \textit{AI Tutor Chat}, and \textit{Ask Human Tutor}. Instructors have the ability to define grading penalties depending on which support tools are used. In this study, accessing the full solution awarded only 10\% credit for participation, while all other features triggered a uniform 25\% deduction, though these settings are fully customizable per assignment.
\end{enumerate}
\end{flushleft}
\end{figure}

\begin{figure}[h]
\centering
\includegraphics[width=0.8\textwidth]{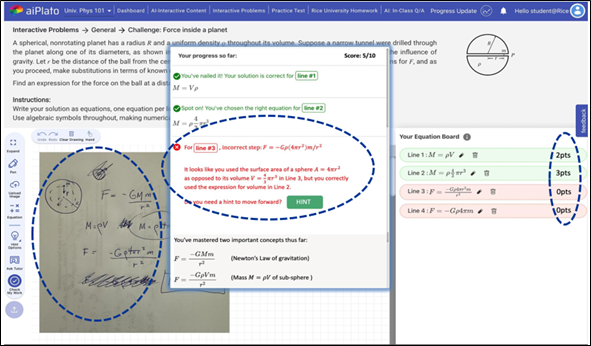}
%\vspace{-1em} % <-- optional if caption is too far
\caption{Handwritten input with real-time feedback and partial credit per line}
\label{Fig. 3}
\begin{flushleft}
  \small Although handwriting input was available, the majority of the students like to type \citep{mogeyStudentsChoicesTyping2012, chwatImpactTechnologyAccess}:  preferred the equation keyboard for speed and reliability. Handwriting recognition issues were common due to doodles, scribbles, and/or cross-outs. See Section \ref{subsec4.3}  for further discussion on student perceptions of aiPlato.   
\end{flushleft}

\end{figure}

\begin{figure}[h]
\centering
\includegraphics[width=\textwidth]{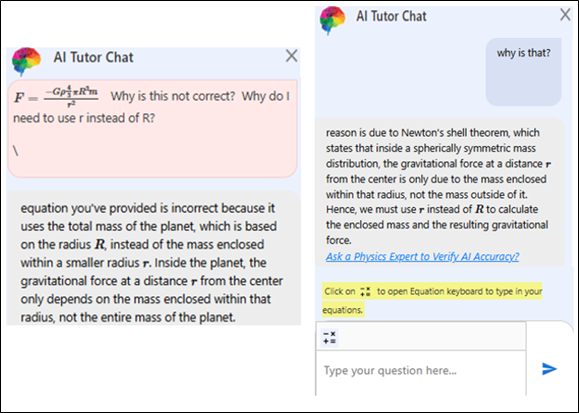}
\vspace{-1em} % <-- optional if caption is too far
\caption{aiPlato AI Tutor Chat explaining conceptual reasoning behind an error}
\label{Fig. 4}
\end{figure}

When students required further clarification, they could initiate a conversation with the built-in \textit{AI Tutor Chat}. Fig. \ref{Fig. 4} shows an interaction in which a student asked why the gravitational force formula should use the enclosed spherical mass at radius r rather than the full mass of the planet R. The AI assistant responded with a conceptual explanation referencing Newton’s shell theorem, helping the student understand the underlying principle rather than just correcting the equation. This tool allowed for asynchronous exploration of reasoning, although students were encouraged to verify AI responses with instructors if needed.

Research shows that almost 83\% of students use email for assignment clarification \citep{weissExplorationEmailCommunication2008}. Many students can become frustrated when professors take too long to respond to their emails, while professors, in turn, may feel overwhelmed by the sheer volume of student queries flooding their inbox \citep{tatumStudentEvaluationsInstructor2021}. From the author’s own experience, tools like \textit{Evaluate My Work} and \textit{AI Tutor Chat} led to a noticeable reduction in the number of student emails related to homework. In prior semesters, the author often received frequent messages asking for clarification on individual steps, problem setup, or whether a particular approach was valid. With aiPlato, many of these routine questions were addressed directly within the platform. Students were able to receive immediate feedback, revise their work, and ask conceptual questions without waiting for an instructor response. Because these tools were integrated into the assignment workspace, the support felt seamless, and in many cases, more accurate and context-sensitive than external platforms like ChatGPT .

\subsection{Student Engagement and Usage Patterns }\label{subsec3.1}

Of the 114 students enrolled in the course, 87 students voluntarily signed the Institutional Review Board (IRB) approved informed consent form, agreeing to have their anonymized interaction data with aiPlato analyzed for research and publication purposes. It is important to note that no students in the class were excluded from using the aiPlato platform or from accessing the extra credit assignments. All students had equal opportunity to benefit from the system’s AI-based feedback and problem-solving tools; only the data from consenting students were used in the following analyses.

\begin{figure}[h]
\centering
\includegraphics[width=0.8\textwidth]{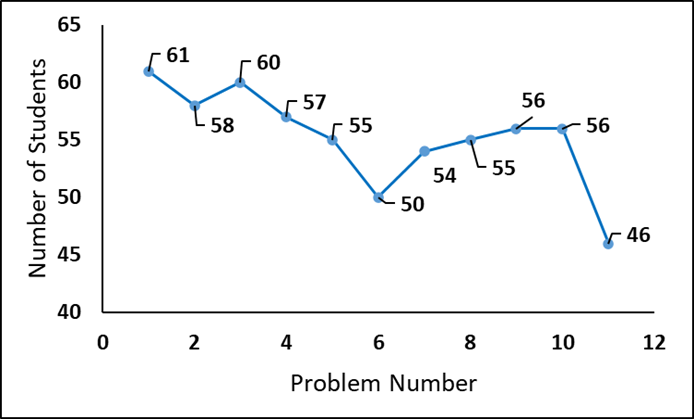}
%\vspace{-1em} % <-- optional if caption is too far
\caption{Number of students attempting each problem in the Extra Credit 4 assignment}
\label{Fig. 5}
\end{figure}

This section presents descriptive patterns in student engagement and tool usage drawn from these 87 students. The instructor had the ability to generate and access detailed usage logs and learning analytics at both the assignment and problem level. Fig. \ref{Fig. 5} shows the number of students attempting each problem within the Extra Credit 4 assignment. While 61 students attempted the first problem, participation gradually decreased across the assignment, with only 46 students attempting the final problem. This trend is consistent with typical patterns of student fatigue or disengagement over assignments. The problem of student attrition is well known \citep{botelhoRefusingTryCharacterizing2019} and such assignment-level analytics, automatically provided by aiPlato, allow instructors to identify potential drop-off points and adjust problem sequencing or pacing in future iterations.

\begin{figure}[h]
\centering
\includegraphics[width=0.8\textwidth]{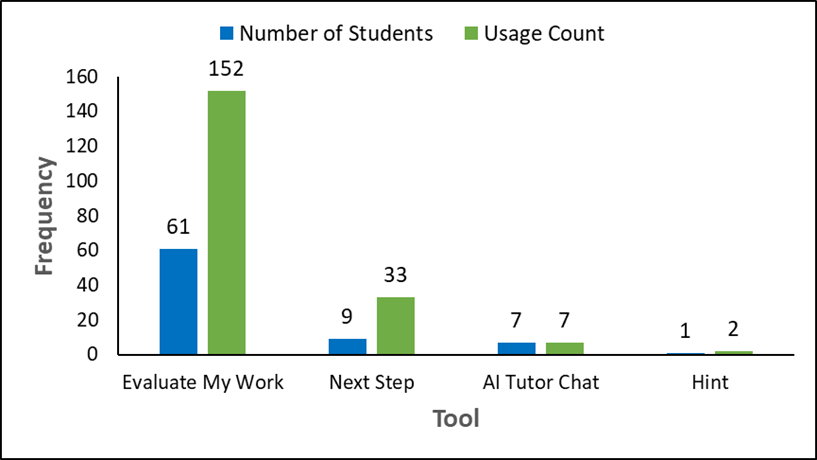}
%\vspace{-1em} % <-- optional if caption is too far
\caption{Frequency of student interaction with AI feedback tools in problem 1 of Extra Credit 4}
\label{Fig. 6}
\end{figure}

Fig. \ref{Fig. 6} displays how students interacted with different AI guidance tools in the first problem of the Extra Credit 4 assignment. All 61 students used the \textit{Evaluate My Work} feature, with a total of 152 uses, indicating that most students engaged in multiple rounds of revision and self-correction. In contrast, more directive tools like Next Step were used by only 9 students, though multiple times (33 total). Tools like AI Tutor Chat and Hint were rarely used. 

These usage patterns support the idea that students engaged in productive struggle. Although this suggests meaningful engagement with the learning process, further research is needed to understand why students prefer certain modes of assistance over others in online homework systems and AITS platforms. Investigating the underlying student decision making processes, including what prompts a student to use \textit{Evaluate My Work} versus more solution revealing tools like \textit{Next Step} or \textit{AI Tutor Chat} (despite similar penalty), remains an open question. A future study involving student interviews and analysis of think aloud protocols while students interacting with the platform would offer valuable insight into students’ cognitive strategies and motivations.

Beyond problem/assignment level statistics, the platform also provides proficiency maps that allow instructors to track student learning across chapters and concepts, both at the individual and class level. At the individual level, the Student Proficiency Map aggregates a student’s performance across all instructional modules and assignments, enabling instructors to identify personal strengths and weaknesses. At the class level, the Course Proficiency Map displays aggregated mastery by topic across the class, helping instructors evaluate which instructional sections may need reinforcement. These visualizations can serve as diagnostic tools to understand whether instructional materials, homework, or in-class interventions are effectively promoting conceptual understanding and skill development.

\begin{figure}[h]
\centering
\includegraphics[width=0.8\textwidth]{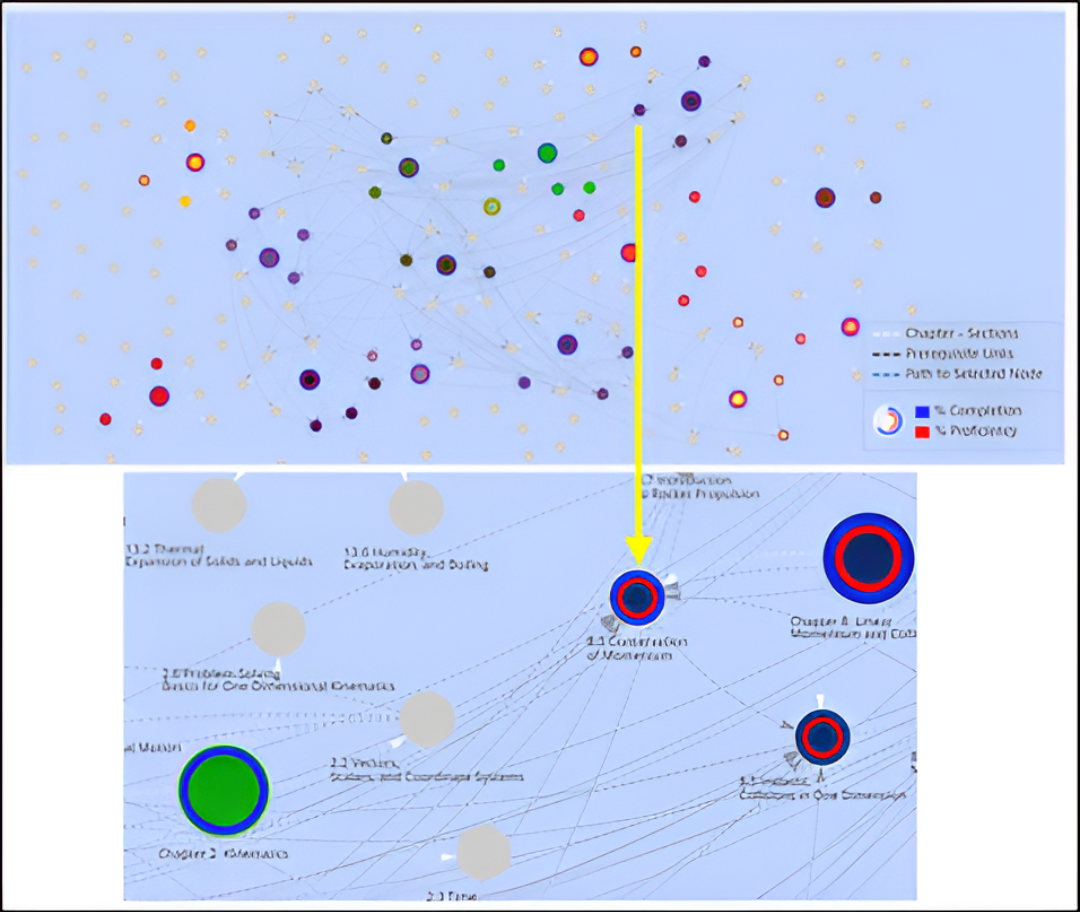}
%\vspace{-1em} % <-- optional if caption is too far
\caption{Course-level proficiency map for PHYS 1443 (Spring 2025) at UT Arlington}
\label{Fig. 7}
\begin{flushleft}
\small
The node for 8.3 Conservation of Momentum (emphasized by the yellow arrow) displays both a red and blue ring, whereas Chapter 2: Kinematics shows only a blue ring, suggesting completion without demonstrated proficiency. Other nodes, like 13.6 Humidity, Evaporation and Boiling, with no rings, were not covered or assessed within aiPlato.
\end{flushleft}
\end{figure}

Fig. \ref{Fig. 7} illustrates the Course Proficiency Map for the PHYS 1443 Mechanics course at the University of Texas at Arlington, Spring 2025, with each node representing a chapter or instructional section. The blue ring indicates the average completion rate and the red ring shows the average proficiency attained by students. The blue outer ring represents the percentage of students who completed problems related to a specific topic, while the red inner ring represents the percentage of students who demonstrated proficiency in that topic. A node with a full blue ring but little or no red ring indicates that most students attempted the problems, but few solved them correctly, highlighting a potential conceptual gap. Conversely, a visible red ring without a significant blue ring suggests that although no problems were directly assigned under that topic, students demonstrated mastery of its underlying concepts in other contexts. aiPlato captures the inherent interconnectedness of physics, allowing proficiency in a given concept to be inferred through successful application in related topics, even when no questions were explicitly assigned under that node.

\section{Spring 2025}\label{sec4}

This section presents the core results from the classroom-based deployment of aiPlato during the Spring 2025 semester of PHYS 1443: Technical Physics I at the University of Texas at Arlington. This was a large-enrollment, calculus-based mechanics course taught entirely on campus with two 80-minute lectures per week. Students attended in-person classes and completed weekly assignments using the traditional ExpertTA platform. In addition to the required homework, students were offered four optional extra-credit assignments on aiPlato (see Section \ref{sec3}).

The goal of this section is to examine whether patterns of student engagement with aiPlato were associated with academic performance, particularly on the cumulative final exam, which constituted 20\% of the course grade. Unlike standard ExpertTA homework, aiPlato allowed students to receive real-time feedback on open-ended problems, revise their work multiple times, and interact with the tools and features described in Section \ref{sec3}. All students enrolled in the course had access to aiPlato; however, participation in the research component of the study was voluntary and conducted under IRB approval.

Because engagement with aiPlato was not randomly assigned, the analyses presented in this section are associational in nature and do not support causal inference. To contextualize observed patterns, we triangulate evidence from multiple sources, including cumulative final exam scores, measures of prior academic performance collected before the introduction of aiPlato, end-of-semester course grades excluding extra-credit contributions, and qualitative survey responses. The first subsection examines the relationship between aiPlato engagement and final exam performance, the second analyzes associations between engagement and final letter grade outcomes, and the third explores student perceptions of aiPlato based on end-of-semester survey data (see Appendix \ref{secA}, \ref{secB}, and \ref{secC} for statistical assumption tests, Appendix \ref{secD} for the survey instrument, and Appendix \ref{secE} for the full dataset).

\subsection{Effect of aiPlato on Final Exam Scores}\label{subsec4.1}

To evaluate whether engagement with aiPlato influenced final exam performance, we first analyzed scores across three levels of platform usage: low, medium, and high engagement. The final exam, worth 20\% of the course grade, and a mixture of multiple choice and free response, was administered in person on May 05, 2025, and assessed all major topics covered throughout the semester, including mechanics, energy, fluids, and thermodynamics. 

Engagement was defined as the combination of a student submitting at least one problem in any of the four aiPlato extra credit assignments (i.e., an “attempt”) and their overall usage of the platform’s feedback features (e.g., \textit{Evaluate My Work}, \textit{AI Tutor Chat}, etc.). A student’s average attempt percentage (calculated across all four aiPlato assignments) served as the primary metric for binning:

\begin{itemize}
  \item Low engagement: Less than 30\% average attempt rate
  \item Medium engagement: Between 30\% and 70\% average attempt rate
  \item High engagement: Greater than 70\% average attempt rate
\end{itemize}

Engagement is associated with many student benefits: academic success, positive development and low drop out rates \citep{fredricksSchoolEngagementPotential2004, fallHighSchoolDropouts2012}. While engagement can be measured across multiple tiers \citep{altuwairqiNewEmotionBased2021}, we categorized students into three categories of low, medium, and high engagement, to balance interpretability with sensitivity to meaningful behavioral differences, while avoiding over-fragmentation given the available sample size. To test the null hypothesis, a one-way Analysis of Variance (ANOVA) was conducted with final exam score as the dependent variable and engagement group (low, medium, high) as the independent factor. The results of the one-way ANOVA and post hoc Tukey Honesty Significant Difference (HSD) tests, which compare all possible pairs of group means, are shown in Table \ref{tab:anova_tukey}. The ANOVA yielded a statistically significant result ($p < 0.05$), indicating that at least one engagement group differed in mean final exam score. Post hoc analysis revealed a statistically significant difference between the low- and high-engagement groups, indicating that higher levels of engagement with aiPlato were associated with higher final exam performance.

While the ANOVA results indicate group-level differences, they do not account for prior academic performance. For example, students with higher engagement may have entered the course with stronger preparation or study habits, making it unclear whether observed differences can be attributed to engagement with aiPlato alone. To address this, we conducted an Analysis of Covariance (ANCOVA) to adjust for measures of prior performance that strictly preceded the introduction of aiPlato. The following covariates were included:

\begin{itemize}
  \item Exam 1 score (administered before introduction of aiPlato to the class)
  \item Pre-aiPlato homework average (homework assignment 0 to 4)
  \item Pre-aiPlato quiz average (Quiz 1 and 2)
\end{itemize}

\begin{table}[h]
\caption{One Way ANOVA and post hoc Tukey HSD Analysis for 87 students: 39 low, 32 med and 16 high engagement level with aiPlato}
\begin{tabular}{@{}p{0.95\textwidth}@{}}
\toprule
\textbf{Null Hypothesis $H_0$:} There is no significant difference in mean final exam scores among students with low, medium, and high engagement with aiPlato. \\
\midrule
\end{tabular}
%--- ANOVA Section (cleaned, tighter fit)
\small
\renewcommand{\arraystretch}{1.05}  % Optional: slightly tighter rows
\resizebox{0.95\textwidth}{!}{%
\begin{tabular}{lccccc}
\multicolumn{6}{l}{\textbf{One Way ANOVA}} \\
\hline
\textbf{Source} & \textbf{Sum of Squares SS} & \textbf{Degrees of Freedom $f$} & \textbf{Mean Square MS} & \textbf{F statistic} & \textbf{p-value} \\
\hline
Treatment & 2,191.39  & 2  & 1,095.70 & 3.94 & 0.0232 \\
Error     & 23,368.27 & 84 & 278.19   &      &        \\
Total     & 25,559.66 & 86 &          &      &        \\
\hline
\end{tabular}
}

\vspace{1em}
%--- Tukey HSD Section
\begin{tabular}{@{}lccc@{}}
\multicolumn{4}{l}{\textbf{Tukey HSD}} \\
\toprule
Pairwise Comparison & Q statistic & p-value & \begin{tabular}[c]{@{}l@{}}Significance\end{tabular} \\
\midrule
Low vs Med   & 1.16  & 0.67  & $p > 0.05$, Not Significant \\
Low vs High  & 3.96  & 0.017 & $p < 0.05$, Significant \\
Med vs High  & 2.93  & 0.1   & $p > 0.05$, Not Significant \\
\botrule
\end{tabular}
%\footnotetext{Source: Analysis results for the effect of aiPlato engagement levels on final exam scores.}
\label{tab:anova_tukey}
\end{table}

The ANCOVA results are summarized in Table \ref{tab:ancova}. After adjusting for prior academic performance, engagement level remained a statistically significant predictor of final exam scores. Among the covariates, Exam 1 score emerged as a strong predictor of end-of-semester performance, consistent with established findings in physics education research \citep{meghjiEarlyDetectionStudent2023, jensenMidtermFirstExamGrades2014, pittmanHowEarlyEarly2021}. In contrast, pre-aiPlato homework and quiz averages were not statistically significant predictors. 

The limited predictive power of these early coursework measures may reflect their restricted scope and timing. Only a small number of assignments (Homework 0--4 and Quizzes 1--2) were completed prior to the introduction of aiPlato, and these covered a relatively narrow portion of the semester’s content. Additionally, relatively few final exam questions were drawn directly from these early topics, further reducing their ability to explain variation in final exam scores.

Taken together, these results indicate that higher levels of engagement with aiPlato were associated with higher final exam performance, even after adjusting for measured prior academic performance. However, because engagement was not randomly assigned, these findings should be interpreted as associational rather than causal. More controlled studies, such as parallel-section or randomized designs, planned as part of the authors’ future work.

\begin{table*}[h]
\caption{ANCOVA for 87 students to account for prior ability}\label{tab:ancova}
\begin{tabular}{@{}>{\centering\arraybackslash}p{2.8cm}cccccc@{}}
\toprule
Source & \begin{tabular}[c]{@{}c@{}}Sum Sq.\\ SS\end{tabular} & \begin{tabular}[c]{@{}c@{}}Df\\ $f$\end{tabular} & \begin{tabular}[c]{@{}c@{}}Mean Sq.\\ MS\end{tabular} & F & $p$ & Significance \\
\midrule
aiPlato Engagement Level   & 1987.96   & 2  & 993.98   & 4.66  & 0.012     & Significant \\
\hline
Exam 1 Score              & 5591.14   & 1  & 5591.14  & 26.2  & $<$0.001  & Significant \\
\hline
Pre-aiPlato Homework Avg  & 0.53      & 1  & 0.53     & 0     & 0.96      & Not Significant \\
\hline
Pre-aiPlato Quiz Avg      & 24.29     & 1  & 24.29    & 0.11  & 0.737     & Not Significant \\
\hline
Error                     & 17283.06  & 81 & 213.37   &       &           &               \\
\botrule
\end{tabular}
%\footnotetext{Source: ANCOVA results accounting for prior ability using exam and pre-aiPlato measures.}
\end{table*}

The mean exam scores for each group were:

\begin{itemize}
  \item Low engagement: 71.53 (n = 39)
  \item Medium engagement: 74.80 (n = 32)
  \item High engagement: 85.39 (n = 16)
\end{itemize}

\begin{figure}[h]
\centering
\includegraphics[width=0.8\textwidth]{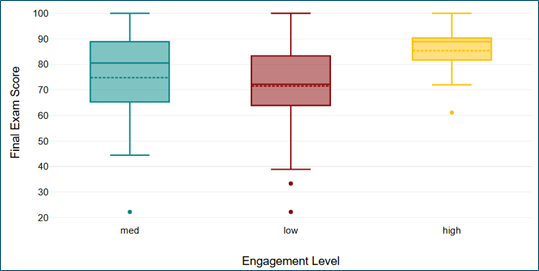}
%\vspace{-1em} % <-- optional if caption is too far
\caption{Box plot showing final exam scores across aiPlato engagement levels. Mean and median showcased by dotted and solid lines respectively}
\label{Fig. 8}
\end{figure}

Figure \ref{Fig. 8} illustrates the distribution of final exam scores across engagement levels. Students in the high-engagement group scored, on average, 13.86 percentage points higher than students in the low-engagement group, after adjusting for prior performance by using covariates that strictly precede the aiPlato rollout (Quizzes 1–2 and Homework 0–4 and Exam 1). This difference was observed following four optional aiPlato assignments and should be interpreted as encouraging in the context of a limited intervention.

\begin{figure}[h]
\centering
\includegraphics[width=0.8\textwidth]{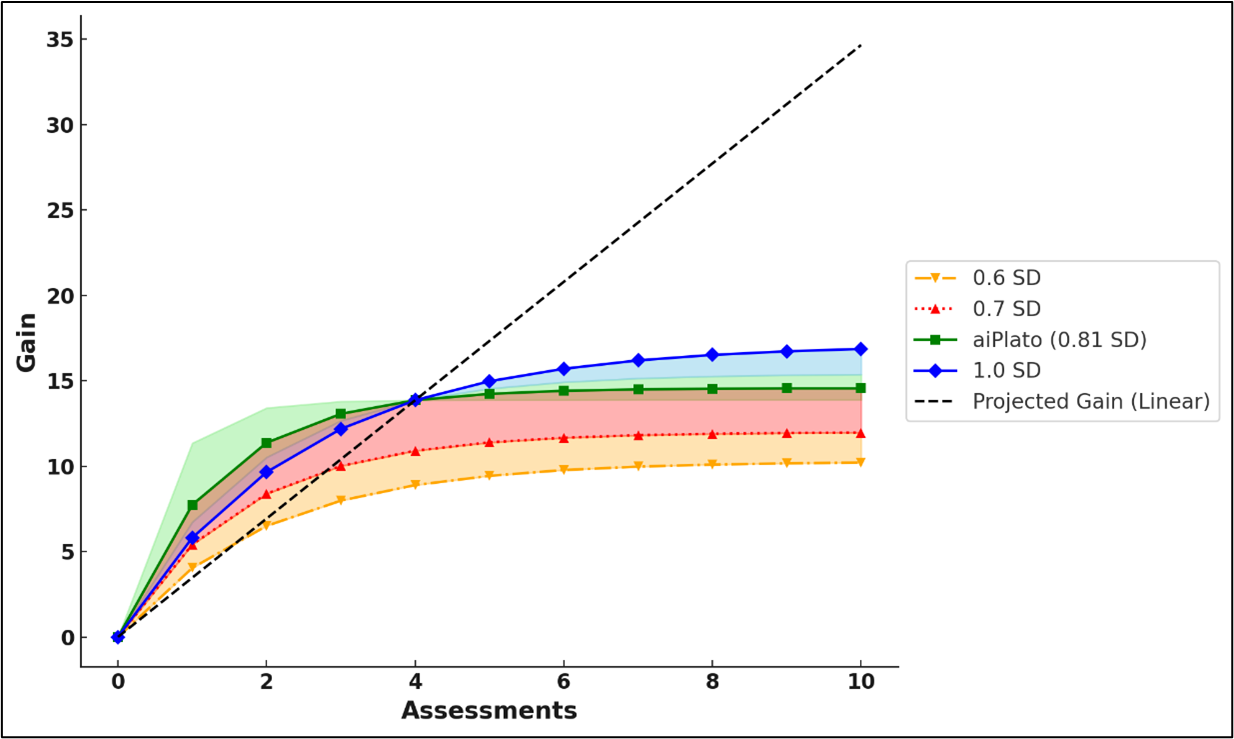}
%\vspace{-1em} % <-- optional if caption is too far
\caption{Learning gains extrapolated under a logistic model for aiPlato compared to benchmark interventions. Shaded regions reflect typical variability in learning effects across different instructional contexts}
\label{Fig. 9}
\begin{flushleft}
\begin{enumerate}
\small \item \textbf{0.6 SD -- 0.7 SD}: Typical of robust interventions like collaborative learning, or deliberate practice \citep{alacapinarEffectCooperativeLearning, KagansFREEArticles}. \\
\small \item \textbf{0.7 SD -- 0.8 SD}: Associated with one-on-one tutoring and intensive formative feedback: Deliberate practice, classroom discussion, scaffolding, etc. \citep{hattieJohnHattieVisible2011}\\
\small \item \textbf{0.8 SD -- 1.0 SD}: Effect sizes in these ranges typically reflect an upper bound that is seldom exceeded except under highly optimized instructional contexts - such as peer instruction or peer-mediated interventions \citep{ozEffectsPeerInstruction2024, haasUnderstandingEffectSize2022, bowman-perrottAcademicBenefitsPeer2013}.
\end{enumerate}
\end{flushleft}
\end{figure}

To explore how observed differences might scale under broader use, we included a hypothetical extrapolation of exam score differences over a full-semester deployment of 10 aiPlato assignments using a logistic growth model that reflects diminishing marginal returns \citep{bettingerNBERWORKINGPAPER, shiPlateauPhenomenonCollege2024}. The observed difference of 13.86 percentage points corresponds to an effect size (Cohen’s $d$) of approximately 0.81 standard deviations over four assignments (see Appendix \ref{secB} for full details).

These benchmark curves, shown in Fig.~\ref{Fig. 9}, are intended as illustrative projections rather than empirical outcomes. A linear extrapolation from four assignments would imply an unrealistically large gain by assignment ten; the logistic model provides a more conservative scenario by explicitly incorporating saturation effects. As such, these projections are presented to contextualize the magnitude of the observed associations and to motivate future controlled studies, rather than to predict instructional impact. Our conclusions remain educationally cautious and comparison studies with a control class over a full semester of aiPlato are within the scope of future work.

\subsection{Final Grade Categorical Analysis}\label{subsec4.2}

This section examines whether student engagement with aiPlato was associated with differences in overall course letter grades, excluding extra credit awarded for aiPlato participation. The goal of this analysis is to characterize the relationship between engagement and overall academic outcomes, while accounting for statistical and sample-size constraints.

Students were grouped into low, medium, and high engagement categories based on their average attempt rate across aiPlato assignments. Final grades were binned as follows: High Grade (A or B) and Low Grade (C or D). Table \ref{tab:grades} summarizes the grade distribution:

\begin{table*}[h]
\caption{Distribution of Final Course Grades (removing effect of extra credit) by aiPlato Engagement Level.}\label{tab:grades}
\begin{tabular}{@{}>{\centering\arraybackslash}p{3.2cm}ccc>{\centering\arraybackslash}p{3.2cm}@{}}
\toprule
Engagement Level & \begin{tabular}[c]{@{}c@{}}High\\ Grade (A/B)\end{tabular} & \begin{tabular}[c]{@{}c@{}}Low\\ Grade (C/D)\end{tabular} & Total & \% High Grade \\
\midrule
Low    & 24 & 15 & 39 & 61.5\% \\
Medium & 25 & 7  & 32 & 78.1\% \\
High   & 15 & 1  & 16 & 93.8\% \\
\botrule
\end{tabular}
%\footnotetext{Source: Classification of student outcomes by aiPlato platform engagement level.}
\end{table*}

To mitigate sparsity of data, letter grades had to be binned into a binary outcome: high performance (A or B) versus lower performance (C or D). This was necessary due to the small number of students in certain categories, particularly the high engagement group, where full A to D granularity resulted in high variance and low expected cell counts, complicating the statistical testing.

\begin{figure}[h]
\centering
\includegraphics[width=0.85\textwidth]{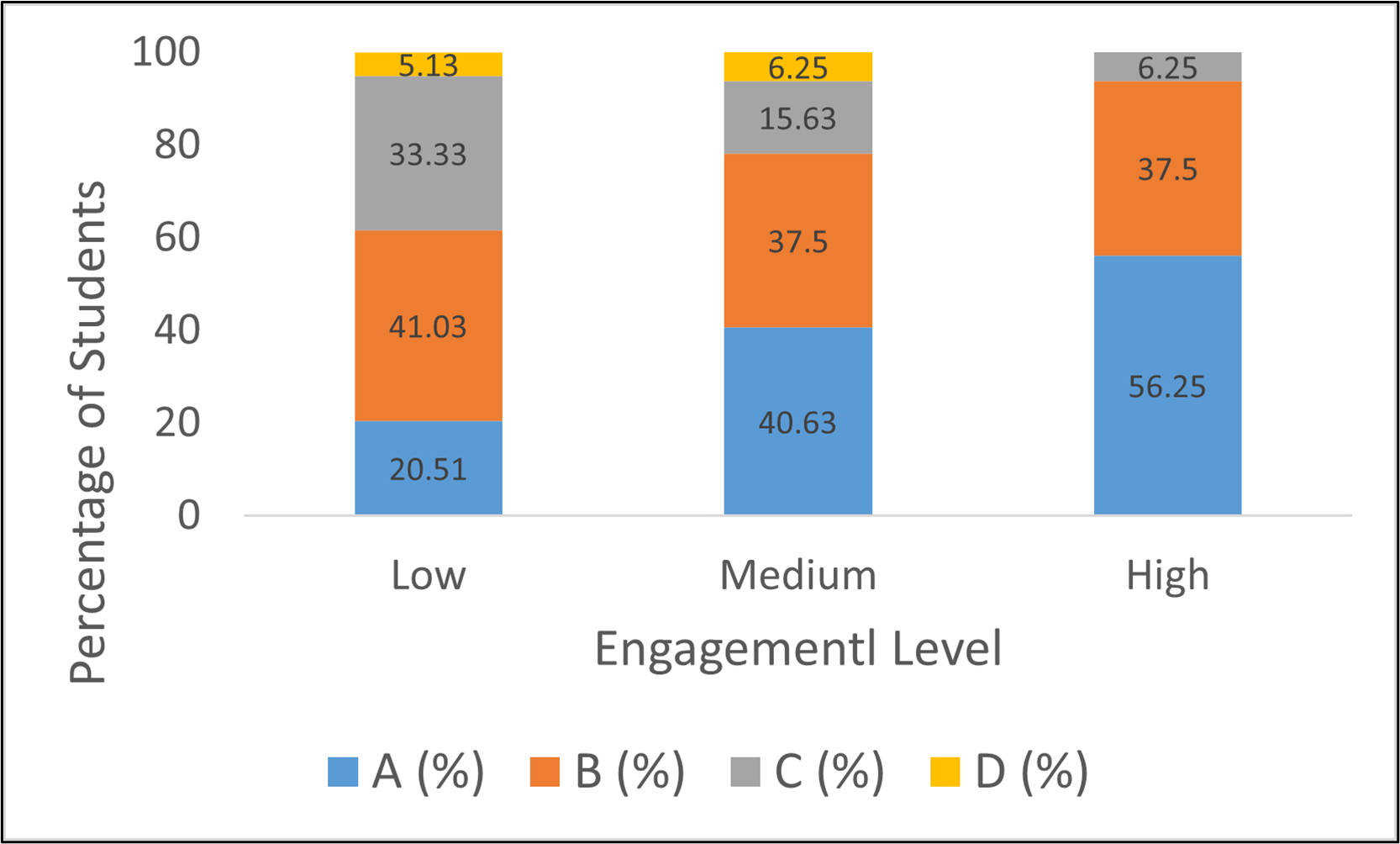}
%\vspace{-1em} % <-- optional if caption is too far
\caption{100\% stacked bar chart showing final course grade distribution (excluding aiPlato extra credit) across engagement levels}
\label{Fig. 10}
\end{figure}

As engagement with aiPlato increased, the proportion of students earning A or B grades rose while the occurrence of C and D grades declined. Notably, none of the high engagement students received a D, and only one received a C (see Fig. \ref{Fig. 10}).

To formally test the association between engagement level and grade outcome, a Fisher’s Exact Test (Freeman–Halton extension for 2×3 tables) was conducted. This test was chosen over the chi-square test due to the small cell counts in the high engagement group, which would have failed the chi-square assumptions.  

The Fisher’s Exact Test yielded an exact two-tailed p-value of 0.036, indicating a statistically significant association between aiPlato engagement level and the final grade category. Students with higher engagement were significantly more likely to receive A or B grades compared to those with lower engagement. 

To further examine this relationship while adjusting for prior academic performance, we conducted a binary logistic regression predicting the likelihood of earning a high grade (A or B, coded as 1) versus a low grade (C or D, coded as 0). Predictors included aiPlato engagement level (categorical: low, medium, high) as well as Exam 1 score, pre-aiPlato quiz average (Quizzes 1–2), and pre-aiPlato homework average (Homework 0–4), consistent with Section \ref{subsec4.1}.

The model correctly classified 82.76\% of students, with a McFadden’s pseudo-$R^2$ of 0.28, indicating a reasonable model fit for exploratory analysis \citep{hemmertLoglikelihoodbasedPseudoR2Logistic2018}. Students in the high-engagement group exhibited a large odds ratio for earning a high grade relative to the low-engagement group; however, this estimate was accompanied by wide confidence intervals and did not reach statistical significance ($p = 0.101$). The lack of statistical significance is likely driven by limited sample size in the high-engagement group, which reduces estimation precision and statistical power.

\begin{table}[h]
\caption{Binary Logistic Regression Key Results}
\label{tab:logistic_regression}
\small
\setlength{\tabcolsep}{4pt}
\begin{tabular}{%
>{\raggedright\arraybackslash}p{3.1cm}  % Predictor
>{\centering\arraybackslash}p{1.1cm}    % B
>{\centering\arraybackslash}p{1.1cm}    % SE
>{\centering\arraybackslash}p{1.1cm}    % z
>{\centering\arraybackslash}p{1.1cm}    % p
>{\centering\arraybackslash}p{1.8cm}    % Odds Ratio
>{\centering\arraybackslash}p{2.1cm}    % 95% CI
}
\toprule
\textbf{Predictor}      & \textbf{B} & \textbf{SE} & \textbf{z} & \textbf{p} & \textbf{Odds Ratio} & \textbf{95\% CI} \\
\midrule
Constant                & -18.55     & 6.32    & 2.94  & .003 & 0      & 0 -- 0              \\
Engagement (medium)     & 0.87       & 0.62    & 1.42  & .156 & 2.40   & 0.72 -- 8.03        \\
Engagement (high)       & 4.12       & 2.51    & 1.64  & .101 & 61.82  & 0.45 -- 8494.59     \\
Exam 1 score            & 0.01       & 0.02    & 0.39  & .694 & 1.01   & 0.98 -- 1.04        \\
Pre-aiPlato HW avg      & 0.16       & 0.06    & 2.47  & .013 & 1.17   & 1.03 -- 1.33        \\
Pre-aiPlato Quiz avg    & 0.04       & 0.01    & 2.66  & .008 & 1.04   & 1.01 -- 1.06        \\
\bottomrule
\end{tabular}
\end{table}

Interestingly, two prior performance measures, pre-aiPlato homework average (p = 0.013) and quiz average (p = 0.008), emerged as statistically significant predictors of final grade category in the logistic regression analysis. This stands in contrast to the earlier findings in Section \ref{subsec4.2}, where Exam 1 score was the strongest predictor of final exam performance. At first glance, this shift may seem contradictory, but it reflects the deeper logic of the course design: the final exam drew heavily from previous exams’ content, making early test performance a good indicator of success on that specific assessment. However, when it comes to the overall course grade, what truly matters is consistent effort across time \citep{riegerPersistenceStudentsAcademic2022}. Homework and quizzes not only formed a larger share of the grade but also rewarded weekly persistence over isolated bursts of performance. So, while doing well on exams may boost your final test score, what ultimately secures an A or B in the class was doing the homework and quizzes. This pattern reinforces a core truth in STEM education: long-term engagement, not one-off performance, is what drives lasting success \citep{bekkeringClassParticipationStudent2021, saqrLongitudinalAssociationEngagement2023}.

While the logistic regression model demonstrated strong overall accuracy and a substantial odds ratio for high aiPlato engagement, its associated p-value narrowly missed conventional significance thresholds. This suggests that although the observed effect may be meaningful, the current sample size, particularly within the high engagement group, limits our ability to draw firm conclusions. Future studies with larger and more balanced samples are needed to rigorously test the statistical significance of aiPlato engagement when controlling covariate data. 

As such, the authors exercise cautious optimism and refrain from making definitive claims about aiPlato's predictive power for overall course grades. For now, we rely on the statistically significant association observed through Fisher’s Exact Test, which affirms a strong relationship between aiPlato engagement level and students’ final grade category, without inferring causality. This relationship merits further investigation in future work.

\subsection{Qualitative Analysis}\label{subsec4.3}

To complement the quantitative results presented in Sections \ref{subsec4.1} and \ref{subsec4.2}, this section explores students’ perceptions of the aiPlato platform through both Likert-scale survey responses and open-ended feedback. The goal was to understand how students experienced the platform's features, how they felt it supported their learning, and whether these subjective experiences aligned with observed performance gains (see appendix \ref{secD} for the full survey).

\begin{figure}[h]
\centering
\includegraphics[width=0.8\textwidth]{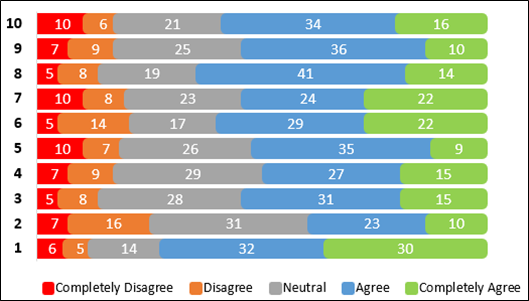}
%\vspace{-1em} % <-- optional if caption is too far
\caption{Diverging stacked bar chart showing student agreement levels with various aspects of the aiPlato platform, based on end-of-semester survey responses}
\label{Fig. 11}
\end{figure}

Post-semester survey responses were collected (n = 87) and the survey included ten Likert-style statements addressing different aspects of the aiPlato experience - from feedback quality to usability and learning impact. Students responded on a five-point scale ranging from Completely Disagree to Completely Agree. The results are visualized in Fig. \ref{Fig. 11} to highlight both positive and negative sentiment across each question. 

Overall, the responses revealed a distinctly positive reception of aiPlato. To further quantify student sentiment, we used the popular Top 2 Box Percentage method. Using this scale, we calculated a composite sentiment score for each survey item, offering a fair and interpretable ranking of features students found most valuable (see appendix \ref{secD} for the full ranked survey item list). This approach not only preserves the directionality of agreement but also emphasizes the intensity of opinion. 

The most agreeable sentiment was:
\begin{itemize}
  \item Question 1: I frequently used the \textit{Evaluate My Work} feature to get step-by step feedback.
\end{itemize}

In contrast, the most disagreeable sentiments was:
\begin{itemize}
  \item Question 2: The \textit{AI Tutor Chat} feature was helpful when I was stuck.
\end{itemize}

To the interest of the authors, Question 6 (Compared to other platforms (e.g., ExpertTA), aiPlato helped make problems more approachable and engaging) and Question 10 (I would recommend the use of aiPlato for future physics assignments and classes) emerged as the third and fourth most positively rated items, respectively. These results provide evidence that students not only perceived aiPlato as more accessible than existing platforms but also expressed a clear desire for its continued use, allowing its potential as a promising instructional tool at large universities and educational institutes. Such feedback directly informs future decisions regarding homework system adoption and supports the authors’ ongoing efforts to improve pedagogical practices through evidence based pilot studies.

The survey also included two open-ended questions inviting students to share what they liked or disliked about aiPlato and suggest areas for improvement. Many students emphasized how aiPlato’s unique features made complex problems more manageable and improved their understanding:

\begin{itemize}
  \item \underline{Step-by-step feedback}: Helped students correct mistakes in real time and understand where their reasoning went wrong
  \item \underline{Multiple feedback options}: Tools like AI Tutor Chat, Hints, and Full Solution gave students flexibility in how they wanted help
  \item \underline{Reduced instructor reliance}: Several students said they emailed the professor far less thanks to the built-in support tools.
  \item \underline{Easier problem comprehension}: Students found aiPlato more intuitive and supportive than traditional homework platforms, especially for multi-step derivations.
  \item \underline{Representative quotes}:
    \begin{itemize}
      \item ``\textit{It provided good feedbacks} [sic] \textit{on incorrect answers and also gave hints for the problems.}''
      \item ``\textit{It evaluated my work and allowed me to keep working on the problem until I got it correct, unlike ExpertTA, which just marked the question wrong and sent me to continue.}''
      \item ``\textit{It was easy to use and gave feedback based on our work. The} Evaluate my work \textit{button was helpful}''
    \end{itemize}
\end{itemize}

Among the negatives, several usability and user interface concerns surfaced in the open responses:

\begin{itemize}
  \item \underline{Handwriting input problems}: Many students struggled with equation recognition and uploads, preferring keyboard entry only.
  \item \underline{Slow/clunky interface}: Described as slow, laggy or over-engineered, particularly on longer assignments or lower-end devices.
  \item \underline{Unreliable AI responses}: Some students reported that AI feedback was incorrect, vague, or misunderstood the context of their question.
  \item \underline{Representative quotes}:
    \begin{itemize}
      \item ``\textit{I personally was not a fan of the AI tutor. I found the hints more helpful than the AI tutor itself. The AI tutor would sometimes get confused or contradicted the feedback I would receive from the} Evaluate my work.''
      \item ``\textit{The} [hand] \textit{writing feature could be completly} [sic] \textit{taken away because it takes to}[o] \textit{long to load and it barely loads properly into the equation section.}''
      \item ``\textit{It was laggy a lot of the times and i}[t] \textit{would forget to submit my solution to the problem.}''
    \end{itemize}
\end{itemize}

Taken together, the Likert ratings and open-ended responses offer a cohesive picture of student sentiment: aiPlato was generally well-received, especially for its detailed feedback and ability to support multi-step reasoning. While some features, such as its user interface and the AI Tutor Chat, were less favorably rated, the overall feedback validates aiPlato’s learning principles and reveals concrete areas for refinement.

In the next and final section, we reflect on the broader implications of this pilot deployment, offer practical recommendations for instructors and developers, and outline directions for future research.

\subsection{Limitations and Future Work}\label{subsec4.4}

A central limitation of classroom-based educational research is the potential for self-selection bias, whereby students who choose to engage with an intervention may differ systematically from those who do not \citep{elstonParticipationBiasSelfselection2021}. In the present study, aiPlato was made available to all students in the course, and extra-credit opportunities were offered uniformly, independent of participation in the research component. However, only a subset of students ($n = 87$) consented to have their data analyzed for research purposes. 

This quasi-experimental design reduces, but does not eliminate, the possibility that observed differences reflect pre-existing differences in motivation, study habits, or academic preparedness. To partially address this concern, our statistical analyses controlled for multiple measures of prior academic performance that preceded the introduction of aiPlato, including Exam~1 scores, early homework averages (Homework~0--4), and quiz averages (Quizzes~1--2). Nevertheless, unmeasured factors such as intrinsic motivation, self-regulation, or time management may still contribute to the observed associations \citep{larzelereInterventionSelectionBias2004}. 

Accordingly, the results of this study should be interpreted as associational rather than causal. Establishing causal effects would require more rigorous designs, such as randomized controlled trials or parallel-section comparisons, which represent a clear direction for future work.

\subsubsection{Engagement Metrics and Behavioral Insights}\label{subsec4.4.2}

Our definition of engagement was based on platform usage logs: problem attempt rates and tool usage frequency. While these offer quantitative insight, they don’t capture how students engaged, whether thoughtfully, strategically, or superficially. To build a more composite engagement index in future studies, methods like think-aloud protocols \citep{wolcottUsingCognitiveInterviews2021}, student interviews, or in-platform reflection prompts \citep{menekseHowDifferentReflection2022} will be essential. These could be paired with usage data to better understand the depth of engagement and how it mediates learning outcomes.

\subsubsection{Statistical Power and Model Complexity}\label{subsec4.4.3}

Although the total sample size was 87 students, certain subgroups, especially the high engagement cohort, were small (n = 16), limiting statistical power. The binary logistic regression included five variables (three covariates and two dummy-coded engagement levels), which may be too complex given the modest sample size. While the odds ratios observed were large and promising, the wide confidence intervals and marginal p-values (e.g., p = 0.101) reflect uncertainty. A larger-scale study with a greater and more balanced sample, particularly in the high engagement group, would enable more stable parameter estimates and clearer conclusions from regression models.

\section{Conclusion}\label{sec5}

This study was guided by three central research questions asked at the end of Section \ref{subsec2.2} : (1) How do students engage with aiPlato’s feedback tools, and what kinds of insights do their interactions provide to instructors about student thinking and learning progress? (2) Is there a measurable relationship between student engagement with aiPlato and their performance on the cumulative final exam? (3) What are students’ perceptions of aiPlato’s usefulness, usability, and learning value, based on end-of-semester survey responses and qualitative feedback? With the completion of our pilot deployment and a triangulated analysis of usage patterns, exam performance, and student feedback, we are now more poised to answer these questions with clarity and evidence.

First, students engaged meaningfully with aiPlato’s feedback tools in ways that suggest genuine and effortful learning. Among the suite of tools, the most frequently used was Evaluate My Work, which students often activated multiple times per problem to test and revise their reasoning. This was followed by more directive tools like Show Next Step, which was used by a smaller group of students but often in bursts, and the AI Tutor Chat, which offered conceptual clarification in response to student queries. Interestingly, the trend across tools showed a clear preference for iterative, formative feedback rather than answer-revealing shortcuts. These patterns, coupled with proficiency visualizations and instructor-facing analytics, gave instructor(s) and the authors insight not just into whether students were getting the right answers, but how they were thinking through each step. aiPlato surfaced conceptual bottlenecks, revealed partial understanding, and allowed instructors to infer mastery across concepts, even when those concepts were not directly tested.

Second, across multiple analyses, higher levels of engagement with aiPlato were associated with stronger academic outcomes, without implying causal attribution. Students in the high-engagement group scored, on average, nearly 14 points higher on the cumulative final exam than students in the low-engagement group, a difference that remained statistically significant after adjusting for prior academic performance. Analysis of overall course grades, excluding extra credit, revealed a similar pattern: students with higher engagement were more likely to earn A or B grades, as indicated by a statistically significant Fisher’s Exact Test. Although logistic regression results suggested large odds ratios for high engagement, the associated uncertainty and marginal p-values highlight the limitations imposed by sample size. Collectively, these findings indicate a consistent association between sustained engagement with aiPlato and academic performance, without establishing causal attribution.

Third, students rated aiPlato favorably across key dimensions of learning value and usability. Survey data showed strong agreement with statements praising the platform’s detailed, stepwise feedback and overall approachability compared to traditional systems like ExpertTA. While students appreciated the learning support features, they also provided constructive feedback about technical issues such as handwriting input glitches, performance issues, user interface and lag. Overall, qualitative responses confirmed that students found aiPlato more supportive, and engaging than other platforms, and many recommended its continued use in future courses.

Taken together, this study positions aiPlato as a promising example of an AI-mediated homework platform that supports open-ended problem solving while generating rich data on student engagement and reasoning. Rather than demonstrating instructional effectiveness in a causal sense, the present work contributes empirical evidence about how such systems are used in authentic classroom settings and how patterns of engagement relate to performance and perception. Future research using larger samples and controlled study designs will be essential to disentangle selection effects, isolate causal mechanisms, and evaluate long-term learning impacts.

This work contributes to ongoing discussions in AI-in-education research about the role of intelligent systems in supporting, rather than replacing, human instruction. The value of AI-enabled platforms may lie less in automation and more in their ability to extend formative feedback and inform instructional decision-making at scale, fostering more reflective and resilient learners \citep{pratamaREVOLUTIONIZINGEDUCATIONHARNESSING2023, chenEnhancingAssessmentPersonalized2023}. As AI systems become more adaptive and better aligned with educational goals, they offer the potential to transform classrooms into environments where every student receives timely and targeted support tailored to their individual learning path \citep{yekolluAIDrivenPersonalizedLearning2024}. The future of AI in education is not about automation, but about equity, and empowerment \citep{katru2025building, shah2023ai}, and that future looks near.

\backmatter

\section*{Declarations}

\bmhead{Conflict of Interest}
The corresponding authors, Atharva Dange and Ramon Lopez, declare that they have no financial or non-financial conflicts of interest related to the subject matter or materials discussed in this manuscript. While aiPlato provided access to their platform for the purposes of this study, there was no financial arrangement or compensation involved. The involvement of aiPlato personnel as co-authors reflects their technical support and platform access, but the corresponding authors maintained full independence in study design, analysis, and interpretation of results. 

%\textbf{\textcolor{red}{aiPlato CONFLICT of INTEREST?}}

\bmhead{Funding}

%\textbf{\textcolor{red}{GRANT FOR PUBLISHING? WHICH GRANT + PUBLISHING COSTS?}} 

\bmhead{Data Availability}
Data sets generated during the current study are available from the corresponding author(s) on reasonable request. Institutional Review Board (IRB) restrictions apply to the availability of these data, and so are not publicly available. .

\bmhead{Author Contributions}
Atharva Dange contributed to the study’s conception and design. Class instruction, data collection, and analysis were performed by Atharva Dange. The final draft of the manuscript was contributed by all authors who commented on previous versions. All authors read and approved the final manuscript.

\begin{appendices}

\newpage
\section{Statistical Assumptions from \ref{subsec4.1}}\label{secA}

To test for normality of data for Section \ref{subsec4.1}, we  conducted the Shapiro-Wilk, Anderson-Darling and the Kolmogorov-Smirnov test, to varying levels of success.

\begin{table*}[h]
\caption{Summary of normality test p-values across engagement groups}\label{tab:normality}
\begin{tabular}{@{}>{\centering\arraybackslash}p{3.0cm}ccc@{}}
\toprule
Engagement Group & \begin{tabular}[c]{@{}c@{}}Shapiro-Wilk\\ p-value\end{tabular} & \begin{tabular}[c]{@{}c@{}}Kolmogorov-\\Smirnov p-value\end{tabular} & \begin{tabular}[c]{@{}c@{}}Anderson-\\Darling p-value\end{tabular} \\
\midrule
Low    & 0.033 & 0.073 & 0.05 \\
Med    & 0.047 & 0.01  & 0.82 \\
High   & 0.20  & 0.077 & 0.161 \\
\botrule
\end{tabular}
%\footnotetext{Source: Normality testing results for final exam score distributions in each engagement group.}
\end{table*}

The Shapiro-Wilk test failed narrowly for both the low and medium engagement groups, whereas the both the Anderson-Darling and Kolmogorov-Smirnov tests did not reject normality for any of the groups. To resolve this inconsistency and further validate the ANOVA results, we also conducted a Kruskal-Wallis (KW) test, a non-parametric alternative that does not assume normality.

\begin{table*}[h]
\caption{Non-parametric Kruskal-Wallis test result comparing final exam scores across aiPlato engagement groups.}\label{tab:chi2}
\begin{tabular}{@{}c c c@{}}
\toprule
$\chi^2$ & Degrees of Freedom (df) & p-value \\
\midrule
8.07 & 2 & 0.018 \\
\botrule
\end{tabular}
\footnotetext{Source: Chi-square test comparing engagement level and grade distribution.}
\end{table*}

KW produced a statistically significant result and since the ANOVA test is more powerful than the KW test and considered robust for moderate violation of the normality assumption, we concluded that the use of ANOVA (and ANCOVA) in Section \ref{subsec4.1} is statistically justified.

\begin{figure}[h]
\centering
\includegraphics[width=\textwidth]{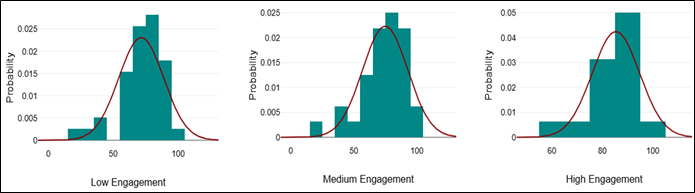}
\vspace{-1em} % <-- optional if caption is too far
\caption{A visual summary of final exam score distributions across engagement groups, illustrating approximate normality in all three groups}
\label{Fig. A1}
\end{figure}

To evaluate the assumption of equal variances across engagement groups, we conducted Levene’s and Brown-Forsythe Test using final exam scores. The test yielded a p-value higher than 0.05 significance, indicating that the variances among the Low, Medium, and High engagement groups are not significantly different. This result supports the assumption of homogeneity of variances, thereby validating the use of ANOVA and ANCOVA, which both require reasonably equal group variances.

\begin{table*}[h]
\caption{Results of Levene’s and Brown-Forsythe tests confirming that variances across engagement groups are approximately equal}\label{tab:levene-bf}
\begin{tabular}{@{}>{\centering\arraybackslash}p{3.3cm}cccc@{}}
\toprule
Test & F-statistic & df 1 & df 2 & p-value \\
\midrule
Levene's Test (Mean)      & 2.77 & 2 & 84 & 0.068 \\
Brown-Forsythe (Median)   & 2.51 & 2 & 84 & 0.087 \\
\botrule
\end{tabular}
\footnotetext{Source: Homogeneity of variance tests for group comparison.}
\end{table*}

To assess the linearity assumption required for ANCOVA, scatter plots were generated showing the relationship between final exam scores and each covariate (Exam 1, pre-aiPlato homework average, and pre-aiPlato quiz average), separated by engagement group. The regression lines were reasonably linear for Exam 1, however the quiz averages and homework average exhibited limited variability and weak predictive trends, suggesting that they may not have been strong indicators of final exam performance - a conclusion that aligns with the results of the ANCOVA model, where pre aiPlato quiz averages and pre aiPlato homework were not significant covariates. 

\begin{figure}[h]
\centering
\includegraphics[width=\textwidth]{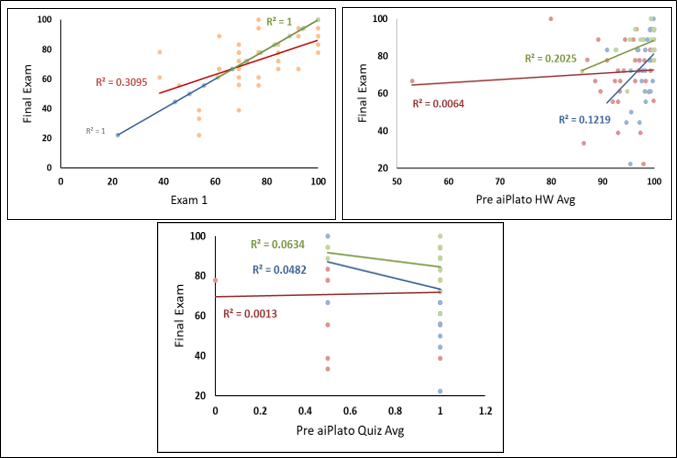}
\vspace{-1em} % <-- optional if caption is too far
\caption{Regression plots showing the relationship between final exam scores and each covariate across engagement groups. Higher R² values indicate a stronger predictive relationship, while lower values (as seen with quiz averages) suggest limited explanatory power}
\label{Fig. A2}
\end{figure}

To test the assumption of homogeneity of regression slopes required for ANCOVA, interaction terms were individually introduced between each covariate (Exam 1, homework average, and quiz average) and the engagement level. In all three regression models, the interaction terms were statistically non-significant, indicating that the relationship between each covariate and the final exam score was consistent across all engagement groups. This supports the assumption that the regression slopes are homogeneous, thereby validating the use of ANCOVA in the main analysis. Furthermore, the lack of significant interactions confirms that none of the covariates behaved differently across low, medium, or high engagement groups, reinforcing the fairness and comparability of adjusted group means.

\begin{table*}[h]
\caption{Summary of interaction terms between each covariate and engagement level. All p-values exceed the 0.05 significance threshold; the assumption of homogeneity is satisfied for all covariates}\label{tab:covariate-interaction}
\begin{tabular}{@{}>{\centering\arraybackslash}p{3.1cm}l c@{}}
\toprule
Covariate & Interaction Term & p-value \\
\midrule
Exam 1                & exam1 $\times$ engagement level               & 0.124 \\
Homework Average      & Pre aiPlato HW avg $\times$ engagement level  & 0.168 \\
Quiz Average          & Pre aiPlato Quiz Avg $\times$ engagement level & 0.288 \\
\botrule
\end{tabular}
\footnotetext{Source: Tests for interaction between covariates and engagement group in ANCOVA model.}
\end{table*}

\newpage
\section{Effect Sizes}\label{secB}

To quantify the strength of aiPlato’s impact on final exam scores, we computed Cohen’s d \citep{becker2000effect},  a standardized measure of effect size commonly used in educational research . It expresses the mean difference between two groups (high vs. low engagement) in units of pooled standard deviation, allowing comparisons across studies regardless of scale.

\begin{equation}
d = \frac{\textit{Mean Difference}}{\textit{Pooled Standard Deviation}} \label{eq:cohensd}
\end{equation}

According to the data and Section \ref{subsec4.2}:

\begin{equation}
d \approx \frac{13.7}{16.9} \approx 0.81\,SD
\end{equation}

To project how learning gains scale with the number of aiPlato assignments, we used a logistic saturation model \citep{meyerPrimerLogisticGrowth, pavlikLogisticKnowledgeTracing2021} that captures diminishing returns in learning:

\begin{equation}
G(n) = L \times \left(1 - e^{-kn}\right)
\end{equation}

Where:

\begin{itemize}
  \item $G(n)=$ Projected gain after $n$ assignments
  \item $L=$ Maximum achievable gain (plateau), where $L = 14.1$

  A ceiling of 14.1 points corresponds to a Cohen’s \textit{d} of 0.85, which is within the upper range of reported learning gains from high-impact interventions in STEM education (e.g., one-on-one tutoring, targeted feedback, peer instruction). Since students already reached 13.7 points of gain after just 4 assignments, it is reasonable to assume that the maximum possible improvement with continued use would be only slightly higher, not dramatically larger. We thus treat 14.1 as an empirically grounded, conservative ceiling.

  \item $k=$ Growth rate, fitted to observed data $\approx 0.89$
  \item $n=$ Number of assignments in the semester $= 10$
\end{itemize}

Thus,

\begin{equation}
G(n) = 14.1 \times \left(1 - e^{-0.89\cdot 10}\right)
\end{equation}

\newpage
\section{Statistical Assumptions from \ref{subsec4.2}}\label{secC}

To evaluate the linearity of the logit assumption for continuous covariates in our logistic regression model, we conducted a post-hoc Box-Tidwell test. This test introduced interaction terms between each predictor and its natural logarithm. The results showed that the linearity assumption was satisfied for the Exam 1 score and the pre-aiPlato HW average but not for the pre-aiPlato quiz average. This violation likely reflects the limitations of the quiz variable, which was based on only two short quizzes administered prior to the aiPlato rollout. With such a narrow window of measurement and limited score variability, the variable may not exhibit a stable linear trend in predicting course grade outcomes. While this limitation warrants cautious interpretation of the quiz coefficient in the logistic regression, it does not undermine the validity of the overall model. Future studies using a larger and more diverse set of pre-intervention quizzes may offer clearer insights into this relationship.

\begingroup
\setlength{\tabcolsep}{5pt} % tighter spacing
\renewcommand{\arraystretch}{1.1} % slightly more row height, optional
\begin{table*}[h]
\caption{Box-Tidwell test results for the linearity of the logit assumption. Only the quiz average term significantly violated the assumption}\label{tab:linearity}
\begin{tabular}{@{}>{\centering\arraybackslash}p{2.5cm} >{\raggedright\arraybackslash}p{3.2cm} c c >{\centering\arraybackslash}p{2cm}@{}}
\toprule
Predictor Variable & Interaction Term & Coefficient $B$ & p-value & \begin{tabular}[c]{@{}c@{}}Linearity\\ Assumption\end{tabular}\\
\midrule
Exam 1 Score               & exam1 score $\times$ $\ln$(exam 1 score)     & 0.00 & 0.425 & Satisfied \\
Pre-aiPlato Homework Avg   & HW avg $\times$ $\ln$(HW\_avg)               & 0.03 & 0.058 & Satisfied \\
Pre-aiPlato Quiz Avg       & quiz\_avg $\times$ $\ln$(quiz\_avg)          & 7.30 & 0.001 & \textbf{Violated} \\
\botrule
\end{tabular}
%\footnotetext{Source: Linearity checks for covariates in ANCOVA analysis using Box-Tidwell transformation.}
\end{table*}
\endgroup

Assumptions of logistic regression were assessed, including linearity of the logit via a Box-Tidwell test. All predictors except one met this assumption. Other diagnostic considerations such as multicollinearity, model fit, and influential cases were not formally tested, as the model was used for exploratory purposes with limited covariates and moderate sample size.

\newpage
\section{Survey Details from \ref{subsec4.3}}\label{secD}

\begin{table*}[h]
\caption{Distribution of student responses across ten aiPlato survey questions using a 5-point Likert scale}\label{tab:likert}
\begin{tabular}{@{}
>{\centering\arraybackslash}p{1.1cm}
>{\centering\arraybackslash}p{1.2cm}
>{\centering\arraybackslash}p{1.1cm}
>{\centering\arraybackslash}p{1.1cm}
>{\centering\arraybackslash}p{1.4cm}
>{\centering\arraybackslash}p{1.9cm}
@{}}
\toprule
Question & \begin{tabular}[c]{@{}c@{}}Completely\\ Agree\end{tabular} & Agree & Neutral & Disagree & \begin{tabular}[c]{@{}c@{}}Completely\\ Disagree\end{tabular} \\
\midrule
Q1  & 30 & 32 & 14 & 5  & 6  \\
Q2  & 10 & 23 & 31 & 16 & 7  \\
Q3  & 15 & 31 & 28 & 8  & 5  \\
Q4  & 15 & 27 & 29 & 9  & 7  \\
Q5  & 9  & 35 & 26 & 7  & 10 \\
Q6  & 22 & 29 & 17 & 14 & 5  \\
Q7  & 22 & 24 & 23 & 8  & 10 \\
Q8  & 14 & 41 & 19 & 8  & 5  \\
Q9  & 10 & 36 & 25 & 9  & 7  \\
Q10 & 16 & 34 & 21 & 6  & 10 \\
\botrule
\end{tabular}
%\footnotetext{Source: Frequency of responses for each survey statement.}
\end{table*}

\begin{table*}[h]
\caption{Survey questions ranked as per the Top Two Box Percentage (T2B\%) analysis (see Section \ref{subsec4.3})}\label{tab:t2b}
\begin{tabular}{@{}>{\centering\arraybackslash}p{0.9cm} >{\centering\arraybackslash}p{1.1cm} >{\centering\arraybackslash}p{1.3cm}@{}}
\toprule
Rank & Question & T2B \% \\
\midrule
1  & Q1  & 71.2  \\
2  & Q8  & 63.2  \\
3  & Q6  & 58.6  \\
4  & Q10 & 57.5  \\
5  & Q3  & 52.9  \\
6  & Q9  & 52.9  \\
7  & Q7  & 52.9  \\
8  & Q5  & 50.6  \\
9  & Q4  & 48.3  \\
10 & Q2  & 37.9  \\
\botrule
\end{tabular}
%\footnotetext{Source: Percentage of respondents rating in the top two categories for each question.}
\end{table*}

\textbf{Survey Instructions for Students}

\textbf{5-point Likert scale:} 1 – Completely disagree \textbar{} 2 – Disagree \textbar{} 3 – Neutral \textbar{} 4 – Agree \textbar{} 5 – Completely agree

\begin{itemize}
  \item Please answer honestly based on your experience using \textit{aiPlato}.
  \item There are no right or wrong answers -- we are looking for your genuine opinions.
  \item Be honest and sincere; your feedback will help improve future versions of the platform.
  \item Thank you for your time and thoughtful responses!
\end{itemize}

\vspace{1em}

\begin{enumerate}
  \item I frequently used the \textit{Evaluate My Work} feature to get step-by-step feedback.
  \item The \textit{AI Tutor Chat} feature was helpful when I was stuck.
  \item I used the AI-generated hints or step-by-step help to guide my problem solving.
  \item I found aiPlato more interactive and engaging than other platforms.
  \item Using aiPlato improved my understanding of physics concepts.
  \item Compared to other platforms (e.g., ExpertTA), aiPlato helped make problems more approachable and engaging.
  \item aiPlato helped me stay motivated and persist longer before giving up, compared to traditional homework platforms.
  \item aiPlato's step-by-step feedback helped identify my mistakes and felt tailored to my specific needs.
  \item Even if it occasionally made mistakes, aiPlato's feedback was generally clear and helpful for learning.
  \item I would recommend the use of aiPlato for future physics assignments and classes.
  \item What did you like or dislike most about aiPlat? (Open ended)
  \item How do you think aiPlato could be improved? (Open ended)
\end{enumerate}

\newpage

\section{Data}\label{secE}

\begin{center}
\hvFloat[nonFloat=true, capPos=top, rotAngle=90, objectPos=c]%
{table}% float type
{% --- BEGIN TABLE BODY ---
\small
\setlength{\tabcolsep}{2pt}
\begin{tabular}{%
    >{\centering\arraybackslash}p{0.5cm}  % Curly brace
    >{\centering\arraybackslash}p{0.9cm}  % Vertical label
    >{\centering\arraybackslash}p{1.4cm}  % Student ID
    |>{\centering\arraybackslash}p{1.3cm} % Exam 1
    |>{\centering\arraybackslash}p{1.5cm} % Final Exam
    |>{\centering\arraybackslash}p{1.5cm} % Pre HW
    |>{\centering\arraybackslash}p{1.5cm} % Pre Quiz
    |>{\centering\arraybackslash}p{2.0cm} % Assignments
    |>{\centering\arraybackslash}p{1.3cm} % aiPlato Avg
    |>{\centering\arraybackslash}p{2.0cm} % Engagement
    |>{\centering\arraybackslash}p{1.1cm} % Help Total
    |>{\centering\arraybackslash}p{2.0cm} % Help on X
    |>{\centering\arraybackslash}p{1.8cm}|} % Grade
\toprule
& & \textbf{Student ID} &
  \textbf{Exam 1 Score} &
  \textbf{Final Exam Score} &
  \textbf{Pre-aiPlato HW Avg} &
  \textbf{Pre-aiPlato Quiz Avg} &
  \textbf{Assign. Attempted} &
  \textbf{aiPlato Avg \%} &
  \textbf{Engagement Level} &
  \textbf{aiPlato Help Total} &
  \textbf{aiPlato Help Used on $x$ Problems} &
  \textbf{Final Grade w/o aiPlato} \\
\cmidrule{1-13}
\multirow{6}{*}{\Huge$\left\{\rule{0pt}{2.5em}\right.$} 
& \multirow{7}{*}{\rotatebox{90}{\textbf{87 Rows}}}
  & 4570 & 76.92 & 66.67 & 99.46  & 100 & 2 & 69.12 & med  & 51 & 40 & B \\
& & \shortstack{$\cdot$\\$\cdot$\\$\cdot$} &  &  &  &  &  &  &  &  &  &  \\
& & 4522 & 84.62 & 83.33 & 98.21  & 100 & 3 & 60.00 & med  & 40 & 39 & A \\
& & 4544 & 38.46 & 78.00 & 86.94  & 100 & 0 & 3.75  & low  &  0 &  3 & C \\
& & \shortstack{$\cdot$\\$\cdot$\\$\cdot$} &  &  &  &  &  &  &  &  &  &  \\
& & \shortstack{$\cdot$\\$\cdot$\\$\cdot$} &  &  &  &  &  &  &  &  &  &  \\
& & 4553 & 69.23 & 66.67 & 52.93  & 100 & 0 & 7     & low  & 15 &  1 & D \\
\bottomrule
\end{tabular}
}% --- END TABLE BODY ---
{Summary of data collected and used for the purposes of this study}
{tab:student_summary_sideways}
\end{center}

\end{appendices}

%%===========================================================================================%%
\newpage
\bibliography{aiPlato_Paper2}% common bib file
%% if required, the content of .bbl file can be included here once bbl is generated
%%\input sn-article.bbl

\end{document}